%% file: main.tex
\documentclass[letterpaper,twocolumn,10pt]{article}
\usepackage{usenix-2020-09}

\usepackage{amsmath}
\usepackage{graphicx}
\usepackage{booktabs}
\usepackage{multirow}
\usepackage{adjustbox}
\usepackage{rotating}
\usepackage{tikz}
\usepackage{xspace}
\usepackage{float}
\usepackage{tabularx}

\newif\ifDEBUG
\DEBUGfalse
\DEBUGtrue

\newif\ifBLINDED
\BLINDEDtrue
\BLINDEDfalse

\newif\ifARXIV
\ARXIVfalse
\ARXIVtrue

\input{misc/typesetting}

\date{}

\title{A Guide to Stakeholder Analysis for Cybersecurity Researchers}

\author{
 {} 
\and
{\rm James C. Davis}\\
Purdue University\\
davisjam@purdue.edu
\and
{\rm Sophie Chen}\\
Carnegie Mellon University\\
scchen@andrew.cmu.edu
\and
{\rm Huiyun Peng}\\
Purdue University\\
peng397@purdue.edu
\and
 {} 
\and
\and
{\rm Paschal C. Amusuo}\\
Purdue University\\
pamusuo@purdue.edu
\and
{\rm Kelechi G. Kalu}\\
Purdue University\\
kalu@purdue.edu
}

\begin{document}

\maketitle

\begin{abstract}
Stakeholder-based ethics analysis is now a formal requirement for submissions to top cybersecurity research venues.
This requirement reflects a growing consensus that cybersecurity researchers must go beyond providing capabilities to anticipating and mitigating the potential harms thereof.  
However, many cybersecurity researchers may be uncertain about how to proceed in an ethics analysis.
In this guide, we provide practical support for that requirement by enumerating stakeholder types and mapping them to common empirical research methods.  
We also offer worked examples to demonstrate how researchers can identify likely stakeholder exposures in real-world projects.  
Our goal is to help research teams meet new ethics mandates with confidence and clarity, not confusion.
\end{abstract}

\section{Introduction}
\label{sec:introduction}
Cybersecurity research is motivated by the goal of improving the safety, privacy, and integrity of computing systems.  
However, the process of conducting and publishing cybersecurity research may itself cause harm.  
Research may expose sensitive information, disrupt services, violate legal or ethical norms, or enable malicious actors by disclosing tools, methods, or vulnerabilities.
Individual researchers~\cite{rogaway2015moral} and academic venues alike have pointed out that ethical considerations in security research are frequently insufficient~\cite{usenix-msg-2025}, perhaps because the authors lack awareness of their work's potential ethical implications.

To mitigate these risks, academic venues increasingly require researchers to engage in ethical analysis.  
For example, the USENIX Security 2026 Call for Papers requires that all submissions include a stakeholder-based ethics analysis on a dedicated supplemental page~\cite{usenix-cfp-2026}.  
This requirement reflects a broader expectation that cybersecurity research must account for its potential harms—not just in methodology, but also in publication.  
A core component of this process is identifying who may be affected by the research.
Before we can reason about effects, we must understand what stakeholders exist, and how they may be impacted.

We recognize that stakeholder-based ethics analysis is a new requirement for the cybersecurity research community, and that not all research teams may be prepared for it.  
We prepared this guide to improve our own research process, and are sharing it to help other research teams get started.  
We specifically focus on the first task: \textit{identifying stakeholders}.  
In~\cref{sec:stakeholder-analysis} we define stakeholder-based ethics analysis and specialize it to cybersecurity. 
\cref{sec:paper-classes} discusses the interaction between research methods and the resultant stakeholders. 
In~\cref{sec:worked-examples}, we provide worked examples of stakeholder identification connected to several papers from our lab.  
Finally,~\cref{sec:discussion} offers guidance on using this paper during research planning or ethics documentation, and discusses topics such as the role of ethical frameworks and the boundaries of responsible research.  

This paper is intended as a practical resource for researchers preparing submissions to USENIX Security and other cybersecurity venues with ethics requirements.  
Researchers may use this work to identify relevant stakeholders in their own studies, understand the kinds of impacts those stakeholders may experience, and articulate their ethical reasoning in a principled and reproducible manner.  
By grounding ethics analysis in examples drawn from typical research practice, we aim to reduce the burden of compliance while increasing the rigor and utility of ethics statements.

\section{Background}
\label{sec:stakeholder-analysis}
We begin by establishing the conceptual foundations of stakeholder-based ethics analysis, situating it within established ethical frameworks and principles that guide computing research. Building on this foundation, we define the notion of a stakeholder in this context and present a structured process for identifying both direct and indirect stakeholders who may be affected by the research process or its outcomes.

\subsection{Ethics Analysis}
\label{sec:concept}

Ethics in security research requires researchers to systematically consider who may be affected by their work and how.  
This approach builds on efforts to bring principled ethical reasoning to computing research.  
The Menlo Report~\cite{menlo-report} articulates four core principles for computing research that were adapted from biomedical ethics: respect for persons, beneficence, justice, and respect for law and public interest~\cite{beauchamp-principles}.  
These principles serve as a baseline for evaluating both research procedures and research outputs.  
Although originally written for information and communication technology (ICT) research, the Menlo principles have since been applied and extended to domains such as adversarial machine learning, vulnerability research, and platform security~\cite{segal-trolley-2023,friedman-2008-value}.  
In these settings, researchers must often weigh tradeoffs between the public good and individual risk, and stakeholder identification becomes a necessary foundation for any such analysis.

Recent work by Segal \etal~\cite{segal-trolley-2023} revisits these principles through the lens of contemporary security research.  
They argue that researchers often face ethical tradeoffs with no clear resolution, such as disclosing a vulnerability versus preventing immediate harm.  
In these cases, considering multiple ethical frameworks—consequentialist~\cite{mill-utilitarianism}, deontological~\cite{kant-metaphysics}, or virtue ethics~\cite{anscombe-modern-moral-philosophy}—can yield a more nuanced analysis.  
\textit{Across all frameworks, however, a common starting point is the identification of stakeholders: those who are potentially harmed or benefited by the research.}

USENIX Security 2026 adopts this principle explicitly.  
Its Call for Papers requires that all submissions include a stakeholder-based ethics analysis or justify an alternative.  
The analysis must describe which stakeholders may be affected by the research process and by the publication of results~\cite{usenix-cfp-2026}.  
The assumption is that ethical research begins with the recognition of who bears the consequences of our work.

\begin{figure}[t]
  \centering
  \includegraphics[width=0.99\linewidth]{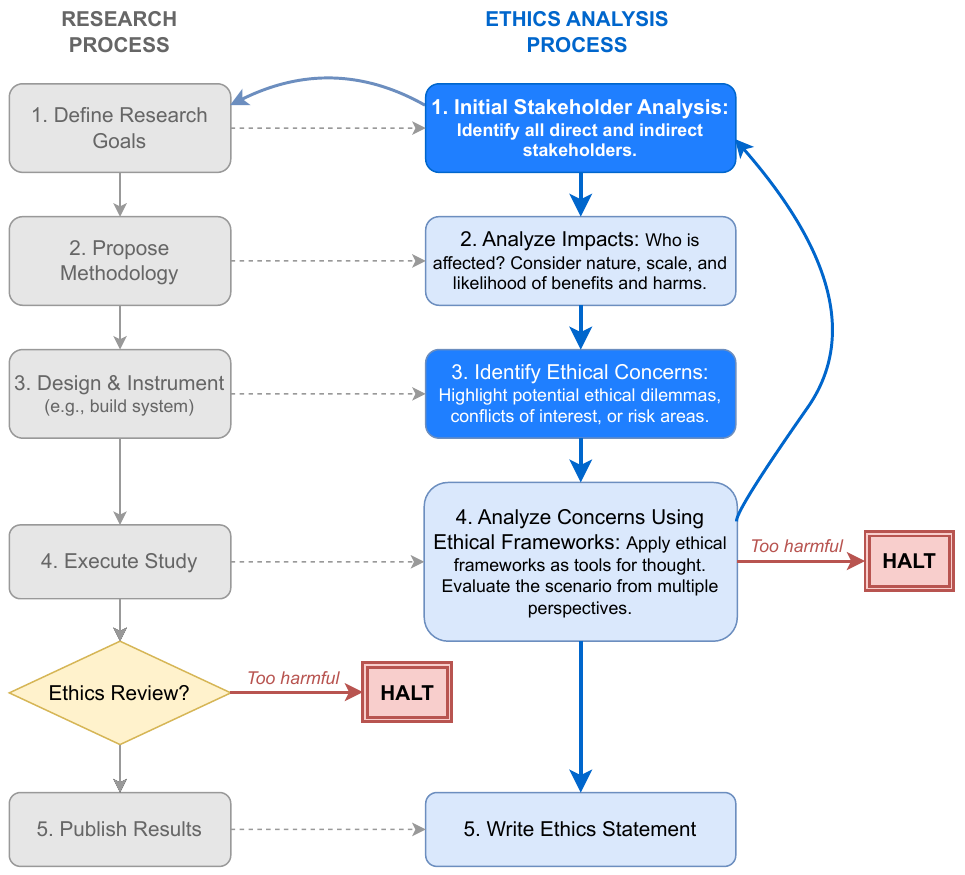}
  \caption{
  Parallel processes of research planning and ethics analysis.
  Arrows represent information flow, with feedback between study execution and mitigation planning.
  Dark blue boxes indicate that stakeholder analysis occurs in (at least) two stages, both during the initial project design (ethics box 1) and during the more detailed design (ethics box 3).
  }
  \label{fig:ethics-flow}
\end{figure}

\Cref{fig:ethics-flow} illustrates how ethics analysis proceeds in parallel with the research process.  
Stakeholder identification begins as soon as research goals are articulated.  
Ethical analysis may influence the proposed methodology, trigger oversight mechanisms, or lead to changes in design, execution, or dissemination. 
When considered early and iteratively, stakeholder analysis can prevent avoidable harms and support more responsible research.

\textit{Beyond serving as a compliance requirement, ethics analysis can also strengthen the quality of security research by clarifying the project's implicit or under-specified requirements}.  
Stakeholder analysis, in particular, forces researchers to consider the expectations, constraints, and vulnerabilities of those affected by the work.
This analysis may reveal new goals, assumptions, or failure modes that an attacker might exploit or a defender might seek to protect. 
This process naturally complements threat modeling by helping researchers define the system boundaries, trust assumptions, and attacker profiles that structure the work~\cite{shostack2014threat}.  
By surfacing these factors early, ethics analysis may prompt adjustments to data collection, modeling choices, evaluation metrics, or publication plans that make the research both more responsible and more robust.  
In this way, ethics reflection can function not as an external constraint, but as a source of epistemic rigor and design clarity.

\subsection{Stakeholder Identification}
\label{sec:identification}

\begin{table*}[t]
  \renewcommand{\arraystretch}{1.2}
  \centering
  \caption{
  Representative stakeholders in cybersecurity research.
  Following the definition in~\cref{sec:identification}, this table distinguishes typical \textit{direct} stakeholders from plausible \textit{indirect} stakeholders.
  This particular example distinguishes between direct and indirect stakeholders by supposing a ``systems'' project that is improving backend systems and incorporating human factors data from the systems' engineering staff.
  }
  \small
  \label{tab:stakeholders}
  \begin{tabular}{m{0.8cm}lp{12.5cm}}
  \toprule
    & \textbf{Stakeholder} & \textbf{Description} \\ \toprule

    \multirow{5}{*}{\centering\rotatebox{90}{\textit{Direct}}} &
    Study participants & Individuals who actively provide data or behavior during user studies, interviews, surveys, etc. \\

    & System operators & People or teams responsible for the systems or services being analyzed or measured (\eg admins of crawled web services). \\

    & Software maintainers & Developers of the software or platforms studied for vulnerabilities, bugs, or compliance. \\

    & Data subjects & Individuals whose personal data appears in datasets, logs, or telemetry used in the study. \\
    
    & The research team & The investigators themselves, who may be exposed to legal, professional, or reputational risk. \\

    \midrule

    \multirow{8}{*}{\centering\rotatebox{90}{\textit{Indirect}}} &
    End users & Users of affected systems who may be harmed by exposure, insecurity, or disruptions. \\

    & Vulnerable populations & Groups disproportionately affected by exploitation or disclosure (\eg activists, minors, marginalized communities). \\
    
    & Adversaries & Actors who might misuse published results to carry out attacks or bypass controls. \\

    & Broader public & Society at large, particularly if trust in infrastructure, institutions, or norms may be impacted. \\

    & Research community & Other researchers who may reuse, replicate, or extend the work and inherit any embedded harms. \\

    & Technology community & Groups interested in security and technology who may implement or adopt the work and inherit any embedded harms. \\

    & Private Institutions & Private organizations, companies, or sponsors whose systems, data, or reputations may be implicated. \\

    & Public Institutions & Public organizations and governments whose systems, data, or reputations may be implicated. \\

    \bottomrule
  \end{tabular}
\end{table*}

\begin{tcolorbox}[title=Definition: Stakeholder, boxsep=2pt, left=2pt, right=2pt, top=2pt, bottom=2pt]
\textbf{A stakeholder} is any person, group, or institution that can affect, or be affected by, a system or its development, operation, or analysis~\cite{795198}.
\textit{Direct} stakeholders interact with, or are explicitly involved in, the research process (\eg as participants or collaborators).
\textit{Indirect} stakeholders may be affected by the consequences of the research, even without direct interaction (\eg through exposure to harms or the use of resulting technologies)~\cite{freeman1984}.
\end{tcolorbox}

\paragraph{General concepts and process.}
Stakeholder identification is the process of enumerating individuals, groups, and institutions who may be impacted by a research project.  
This includes both ``direct stakeholders''—such as study participants or software maintainers—and ``indirect stakeholders'', such as users of affected systems, vulnerable populations, or the broader public.  
The distinction reflects the length and complexity of the cause/effect pathway between research and its stakeholders~\cite{leveson2011engineering}:
  direct stakeholders are typically affected through short, observable links to the research process or artifacts,
  while indirect stakeholders may be impacted through more diffuse or downstream effects, including the adoption, misuse, or unintended consequences of research outputs.

In requirements engineering, stakeholder identification is a foundational step for eliciting system goals, constraints, and assumptions~\cite{795198, sommerville1997guide}.
Stakeholders are often grouped into classes such as \textit{users}, \textit{developers}, \textit{regulators}, and \textit{operators}, with attention to their influence, interests, and modes of interaction~\cite{sommerville1997viewpoints}.  
This framing usefully extends to research ethics: a stakeholder is not merely a data point or source of input, but an actor with values, expectations, and potential to be harmed or helped by the research process or its outcomes. 

\paragraph{Specialization to Security Research.}
Security research is broadly concerned with understanding, analyzing, and improving the trustworthiness of systems, infrastructure, and users.  
The U.S. National Institute of Standards and Technology (NIST) defines cybersecurity as ``the process of protecting information by preventing, detecting, and responding to attacks''~\cite{nist-glossary}.  
The 2025 Calls for Papers for IEEE Symposium on Security and Privacy~\cite{sp-cfp-2025}, USENIX Security~\cite{usenix-cfp-2025}, NDSS~\cite{ndss-cfp-2025}, and ACM CCS~\cite{ccs-cfp-2025} all emphasize novelty, rigor, and impact across a diverse set of technical and human-centered topics.  

The global reach and leverage of computing systems mean that advances can affect people, organizations, and environments far beyond the original research context. Security and IT research often must account for adversarial stakeholders—attackers, competitors, deceivers—who may misuse research outputs or be targeted by defensive tools. These properties blur the boundary between direct and indirect stakeholders, creating distinctive challenges in security research to map activities to affected parties:

\begin{itemize}[leftmargin=*]
    \item \textit{Scale and indirectness:}  
    Security research can produce effects far beyond its original scope, as impacts propagate globally through interconnected systems—often with delayed, indirect, or hard-to-trace consequences.  
    For example, a dataset study may not involve users directly, but could still expose them to privacy harms via re-identification~\cite{narayanan2008robust}.  
    These indirect effects challenge traditional notions of informed consent and risk assessment~\cite{metcalf2016ethics}.

    \item \textit{Dual-use potential:}  
    Research outputs, such as tools, datasets, and attack techniques, can be used both to defend and to harm.  
    A novel exploit, even if disclosed responsibly, may inspire malicious actors before defenses are deployed~\cite{brumley2008automatic}.  
    The longstanding tension between openness and control in dual-use publication is a central concern in cybersecurity~\cite{riebe2019dual}.

    \item \textit{Adversarial response:}  
    Unlike many domains, security research exists in an adversarial context, where attackers may adapt strategically to research disclosures.  
    Published defenses may provoke bypass techniques or trigger new attack variants~\cite{mutlu2019rowhammer}.  
    Security researchers must therefore anticipate not only technical outcomes, but also adversary reactions --- an expectation uncommon in other scientific disciplines.

    \item \textit{Ambiguity in ethical responsibility:}  
    The indirect and adversarial nature of security research can obscure who might be harmed, and who bears responsibility for preventing or mitigating that harm.  
    Attribution is especially difficult: harms may result from chains of actions involving researchers, developers, deployers, and adversaries, with no clear line of causality or control.  
    For example, an insecure design pattern documented for defensive awareness might later be incorporated into offensive malware by an unrelated actor.  
    These ambiguities complicate the mapping from research decisions to ethical duties~\cite{friedman-2008-value}.
\end{itemize}




~\Cref{tab:stakeholders} presents a canonical set of stakeholder categories relevant to common types of cybersecurity research.  
We distinguish between direct and indirect stakeholders, and include brief descriptions of each.  
This list may be extended or refined depending on the project context, but it provides a baseline for constructing the ethics section required by the USENIX Security 2026 CFP.
In~\cref{sec:worked-examples}, we will specialize these categories for example research scenarios.

\section{Research Methods \& Stakeholder Exposure}
\label{sec:paper-classes}


\begin{table*}[t]
  \renewcommand{\arraystretch}{1.4}
  \centering
  \caption{
    Cybersecurity research methods grouped by examples of direct stakeholders.
    Method categories are derived from the SIGSOFT Empirical Standards~\cite{sigsoft-standards}.
    Citations in \textbf{bold} are included in the analysis in~\cref{sec:worked-examples}.
  }
  \resizebox{\textwidth}{!}{%
    \begin{tabular}{lp{6cm}p{6cm}p{3cm}}
      \toprule
      \textbf{Method(s)} & \textbf{Description} & \textbf{Typical Direct Stakeholders} & \textbf{Example Papers} \\
      \toprule

      \textit{Controlled Experiments, Surveys, Interviews} &
      Designed interaction with human participants to study behavior, decision-making, or perception. &
      Study participants  &
      \textbf{\cite{kalu2024industry}},~\cite{285411} \\

      \textit{Case Study, Action Research} &
      In-depth investigation or intervention in a real-world system or organization. &
      Internal developers, administrators, system maintainers. &
     ~\cite{airtag},~\cite{dalle} \\

      \textit{Repository Mining, Corpus Analysis} &
      Analysis of public artifacts such as source code, commits, vulnerability disclosures, or usage telemetry. &
      OSS maintainers, contributors, downstream users. &
      \textbf{\cite{10298483}},~\cite{catch22} \\

      \textit{Tool Evaluation, Benchmarking} &
      Evaluation of tools or techniques against benchmarks to assess effectiveness, efficiency, or coverage. &
      Tool developers, analysts, affected system operators. &
      \textbf{\cite{amusuo2025unit}},~\cite{298150},~\cite{10646663} \\

      \textit{Simulation, Optimization} &
      Modeling or optimization of system behavior under constraints or attack scenarios. &
      Simulated users, designers of real-world systems, policymakers. &
      \textbf{\cite{11029801}},~\cite{298186} \\

      \textit{Longitudinal Studies, Meta-Science} &
      Empirical analyses of trends over time or research practices across populations of studies. &
      Research community, prior authors, funding agencies. &
     ~\cite{285379},~\cite{11023368},~\cite{11023489} \\

      \textit{Systematic Review, Replication} &
      Reproduction or synthesis of published results to assess validity, generalizability, or evidence strength. &
      Original study authors, readers of syntheses, methodology developers. &
     ~\cite{10646648},~\cite{zksnarks} \\

      \bottomrule
    \end{tabular}
  }
  \label{tab:research-methods}
\end{table*}

To interpret the stakeholder categories in~\cref{tab:stakeholders}, we consider how \emph{problem domains} and \emph{research methods} each shape ethical risk in security studies.
Both dimensions influence stakeholder exposure, but in different ways.
For example, studies on password reuse directly implicate end-users and service providers, whereas protocol-level vulnerability research more directly affects software maintainers and operators.
Research methods, in turn, introduce distinct forms of exposure, especially for direct stakeholders.
A quantitative analysis of software packages and a qualitative interview study may both examine supply chain security, but in different ways. The former may reveal vulnerabilities in widely used packages, increasing opportunities for exploitation, while the latter risks compromising participants’ privacy or reputation if insider knowledge about development practices is disclosed.
Thus, problem domains determine \emph{who} is exposed, whereas research methods determine \emph{how} those risks materialize.

To support consistent stakeholder identification across these various types of research, we organize common security research methods into a set of clusters derived from the SIGSOFT Empirical Standards~\cite{sigsoft-standards}.  
~\Cref{tab:research-methods} presents representative groupings of methods that interact with similar kinds of stakeholders. We also provide example security papers from each group, although note that a single work may fall into multiple groups.

\section{Example Studies and Stakeholder Analysis}
\label{sec:worked-examples}

To illustrate how the method–stakeholder framework in~\cref{sec:paper-classes} can be applied in practice, this section presents worked examples of stakeholder identification for representative security research studies.
\cref{tab:example-studies} gives a summary.  
Each example includes a concise project description, identifies the research method(s) and problem domain, and lists the plausible direct and indirect stakeholders.  
Where appropriate, we also include commentary on unusual exposures or particularly salient ethical concerns.
To mitigate concerns about bias or blame, we only discuss papers whose authors are represented on the author list of the present guide.
We selected papers from a range of research methods.

These examples are intended to help research teams anticipate the kinds of stakeholder relationships they may encounter in their own work.  
As described in~\cref{sec:stakeholder-analysis}, stakeholder identification is a precursor to further ethical analysis using frameworks such as deontological, consequentialist, or virtue-based reasoning (cf.~\cref{sec:discussion-ethics}).  
We do not aim to perform full ethics reviews here, but instead to concretely demonstrate the process of naming and describing stakeholder groups.

\begin{table*}[ht]
\centering
\footnotesize
\caption{A summary of the stakeholder analysis examples, including example background and resulting stakeholders}
\label{tab:example-studies}
\begin{tabularx}{\textwidth}{c X X *{2}{X}}
\toprule
\textbf{Example} & \textbf{Context} & \textbf{Method} & \multicolumn{2}{c}{\textbf{Stakeholders}} \\
\cmidrule(lr){4-5}
& & & \textit{Direct} & \textit{Indirect} \\
\midrule
A (\cref{sec:example-a}) &
Vulnerability detection in embedded network stacks (ENS), whose flaws enable remote exploitation. &
Corpus analysis of CVEs; tool design/evaluation on real ENSs &
Software maintainers; research team &
System operators; adversaries; end users; broader public \\

B (\cref{sec:example-b}) &
Organizations adopt bounded model checking but their methods vary and are error-prone, missing vulnerabilities &
Process evaluation; corpus analysis of unit proofs for embedded OSs &
Software maintainers; research team &
System operators; adversaries; end users; broader public \\

C (\cref{sec:example-c}) &
Software signing adoption is uneven despite mandates, with a gap in understanding organizational challenges &
Semi-structured interviews with 18 practitioners from 13 organizations &
Interview participants; their organizations; research team; adversaries &
Other software producers; standards bodies; software consumers; broader public \\

D (\cref{sec:example-d}) &
Modern apps rely on dependencies with reachable vulnerabilities, which can be mitigated with runtime defenses &
Corpus analysis of library vulnerabilities; Zero-Trust Dependencies tool evaluation &
System operators; software maintainers; research team; adversaries &
Java maintainers; end users; broader public \\
\bottomrule
\end{tabularx}
\end{table*}

\subsection{Example A: Software Validation Project}
\label{sec:example-a}


This example represents papers concerned with validating the security of software.
The following analysis is on the paper ``\textit{Systematically Detecting Packet Validation Vulnerabilities in Embedded Network Stacks}'', which appeared at ASE 2023~\cite{10298483}. 
Papers with similar analyses may include~\cite{287290} and~\cite{10646755}. 


\subsubsection{Study Overview}

Embedded Network Stacks connect critical embedded systems to external networks. As a result, vulnerabilities in ENS can be remotely exploited to cause denial of service, arbitrary code execution, or physical-world harm. Prior dynamic analysis based approaches relied on non-deterministic mutations and provided no security guarantees. This work aims to provide a more systematic dynamic analysis framework to uncover security vulnerabilities in these critical components.

\paragraph{\textbf{Problem Domain}}
Detection of security vulnerabilities in embedded network stacks.

\paragraph{\textbf{Methods}}
Two methods were applied:
    \begin{itemize}
        \item \emph{Corpus Analysis}: Analyzed 61 reported vulnerabilities (CVEs) across six embedded network stacks.
        \item \emph{Tool Design and Evaluation}: Designed and evaluated the effectiveness and performance of EmNetTest, a novel systematic testing framework, on real embedded network stacks.
    \end{itemize}

\subsubsection{Stakeholder Identification}

\paragraph{Direct Stakeholders}
\begin{itemize}
    \item \textit{Software maintainers}: Developers and maintainers of ENS studied in the paper (\eg FreeRTOS, Contiki-ng, lwIP, PicoTCP).
    \item \textit{The research team}: Authors of the work, who may face legal, professional, or reputational risks associated with vulnerability discovery and disclosure.
    \item \textit{Adversaries}: Malicious actors who could either exploit disclosed vulnerabilities before they are patched, or use the provided security tool to discover new vulnerabilities in other software products.
\end{itemize}

\paragraph{Indirect Stakeholders}
\begin{itemize}
    \item \textit{System operators}: Organizations or engineers integrating ENSs into their products (\eg, IoT device vendors, integrators).
    \item \textit{End users}: Individuals using products containing vulnerable ENSs, who could be affected by service disruption or compromise.
    \item \textit{Broader public}: Society at large, where IoT failures may erode trust in connected technologies or disrupt essential services.
\end{itemize}

\subsubsection{Ethical Considerations}

\begin{itemize}
    \item \textit{Risk of exploiting new vulnerabilities}: Publicly providing descriptions of new vulnerabilities could facilitate exploitation if accessed before affected systems are patched.
    \item \textit{Risk of exploiting other software of similar characteristics}: Adversaries could use the tool designed in this paper to discover and exploit zero-day vulnerabilities in other network stacks or embedded software.
\end{itemize}

\subsection{Example B: Empirical Software Security Study}
\label{sec:example-b}

This example represents papers concerned with empirically measuring the security of software and the effectiveness of security tools and techniques. 
The following analysis is on the paper ``\textit{Do Unit Proofs Work? An Empirical Study of Compositional Bounded Model Checking for Memory Safety Verification}'', which will appear at ICSE 2026~\cite{amusuo2025unit}.
Papers with similar analyses may include~\cite{298202} and~\cite{298148}.

\subsubsection{Study Overview}

Organizations are increasingly adopting bounded model checking to verify the memory safety of real software. However, their methods for creating these ``unit proofs'' vary and are prone to errors. 
This increases the cost of the process and lead to missed security vulnerabilities.
The goal of this work is to provide an empirical basis for the process to inform other organizations of the costs and benefits.

\paragraph{\textbf{Problem Domain}}
Practical compositional bounded model checking of real software.

\paragraph{\textbf{Methods}}
This study employed two methods:
\begin{itemize}
    \item \emph{Tool/Technique Evaluation}: Evaluated the effectiveness and the cost of systematic unit proofing for memory safety verification.
    \item \emph{Corpus Analysis}: Analyzed the characteristics of the unit proofs used to verify real embedded operating systems.
  \end{itemize}

\subsubsection{Stakeholder Identification}

\paragraph{Direct Stakeholders}
\begin{itemize}
  \item \textit{Software maintainers} — Maintainers of the studied embedded software (Zephyr, Contiki-ng, RIOT-OS, FreeRTOS) and bounded model checking tools (CBMC).
  \item \textit{The research team} — Authors conducting the study.
\end{itemize}

\paragraph{Indirect Stakeholders}
\begin{itemize}
    \item \textit{System operators}: Organizations or engineers integrating ENSs into their products (\eg, IoT device vendors, integrators).
    \item \textit{Adversaries}: Malicious actors who could exploit disclosed vulnerabilities, or be hindered if mitigations are widely deployed.
    \item \textit{End users}: Individuals using products containing vulnerable ENSs, who could be affected by service disruption or compromise.
    \item \textit{Broader public}: Society at large, where IoT failures may erode trust in connected technologies or disrupt essential services.
\end{itemize}

\subsubsection{Ethical Considerations}

\begin{itemize}
    \item \textit{Risk of exploiting new vulnerabilities}: Publicly providing descriptions of new vulnerabilities could facilitate exploitation if accessed before affected systems are patched.
    \item \textit{Risk of exploiting other software of similar characteristics}: Adversaries could use the tool designed in this paper to discover and exploit zero-day vulnerabilities in other network stacks or embedded software.
    \item \textit{Risk of downtime during patching:} Recreated defects and unit-proof results signal patching and deployment risk. Operators may face downtime risks and must plan for timely updates, regression testing, and staged rollouts to avoid service disruption while addressing memory-safety issues.
  
\end{itemize}

\subsection{Example C: Qualitative Study of Organizational Security Practices}
\label{sec:example-c}

This example encompasses studies that investigate and measure security practices and adoption by organizations.
The following analysis is on the paper ``\textit{An Industry Interview Study of Software Signing for Supply Chain Security}'', which was published in USENIX 2025~\cite{kalu2024industry}.
Papers with similar analyses may include~\cite{285517} and~\cite{11023474}.

\subsubsection{Study Overview}

Industry and regulatory bodies increasingly mandate practices like software signing, yet adoption in practice remains uneven and unclear. Existing research has largely focused on technical measurements of signing prevalence, leaving a limited understanding of organizational challenges and practitioner perspectives. This paper addresses that gap by situating signing within broader industry trends, regulatory pressures, and real-world production workflows.

\paragraph{\textbf{Problem Domain}}
This work examines the industrial adoption of Security practices, specifically software supply chain security practices, focusing on the role and challenges of software signing in ensuring provenance and integrity of artifacts.

\paragraph{\textbf{Methods}}
\emph{Interviews}: The study employed a semi-structured qualitative interview instrument with 18 senior security practitioners from 13 organizations. Responses were analyzed using thematic and framework analysis (with the Software Supply Chain Factory Model as a reference) to examine four concerns in software signing adoption — real-world practices, implementation challenges, perceived importance, and the influence of standards, regulations, and security incidents.

\subsubsection{Stakeholder Identification}
\paragraph{\textbf{Direct Stakeholders}}
\begin{itemize}
    \item \textit{Interview Participants}: Practitioners who take part in the study may be at risk if their identities are not properly anonymized, as their participation could unintentionally reveal organizational practices. This exposure may lead to reputational or professional risks if flaws or poor practices are highlighted in the research.   
    \item \textit{Organizations of participating practitioners}: Organizations whose employees participated in this study may face reputational or regulatory scrutiny, or even security risks to their signing infrastructure, if weaknesses in their policies or practices are revealed. These risks are heightened if the identities of participating organizations are not properly anonymized.
    
    \item \textit{The research team}: Responsible for accurate representation and avoiding overgeneralization. They must balance transparency with the risk that exposing gaps could inadvertently harm the organizations studied or aid adversaries. In addition, they face potential legal implications if participant identities or organizational details are inadvertently disclosed, violating confidentiality agreements or data protection regulations.  
    \item \textit{Adversaries}: Could exploit weak or absent signing, and may benefit from publicized research findings if disclosures are not carefully managed. Risks increase if the identities of participating organizations or practitioners are revealed, providing attackers with more direct targets. 
\end{itemize}

\paragraph{\textbf{Indirect Stakeholders}}
\begin{itemize}
    \item \textit{Other software producing-organizations adopting signing}: Beyond the companies directly participating in this study, other organizations that rely on software signing may also face indirect concerns if the research highlights weaknesses in current implementations. Such findings could expose gaps in industry practices, leading to reputational harm, regulatory pressure, or increased scrutiny of their signing infrastructures.

    \item \textit{Standards organizations}: Bodies that produce guidelines for software signing may be indirectly affected if the research exposes gaps or ambiguities in existing standards. Such findings could challenge their credibility, but also create pressure to revise or strengthen their recommendations.

    \item \textit{Software consumers}: Individuals and organizations relying on signed software products may lose confidence in the trustworthiness of signing mechanisms if research findings highlight serious flaws. While such disclosures could ultimately improve long-term security, they also risk short-term confusion or distrust among users.

    \item \textit{Broader public}: Broader society may be affected if research reveals systemic weaknesses in software signing that undermine trust in critical digital infrastructure. Public confidence in software supply chain security could be eroded, potentially discouraging the adoption of secure technologies or fueling fear around the safety of connected products.
\end{itemize}

\subsubsection{Ethical Considerations}

\begin{itemize}
    \item \textit{Risk of reputational harm}: Publicly highlighting organizational shortcomings in signing practices could damage trust in individual companies or entire sectors if anonymity is not carefully maintained.
    \item \textit{Risk of aiding adversaries}: Detailed descriptions of weak or absent signing may be misused by attackers to identify and exploit unprotected software supply chains.
    \item \textit{Risk to participants}: Interviewees may face professional or organizational consequences if their responses are linked back to them, raising concerns about privacy and proper anonymization.
    \item \textit{Regulatory and compliance exposure}: Findings could increase external scrutiny on organizations, potentially triggering audits, penalties, or stricter mandates if deficiencies are publicized.
    \item \textit{Power dynamics}: Practitioners who participated in interviews may not have decision-making power over signing adoption, yet their responses could still expose organizational vulnerabilities or poor practices outside their control.
    \item \textit{Societal trust}: Revealing widespread issues in signing implementations could reduce public confidence in software ecosystems and critical infrastructure that rely on them, even as the research aims to improve overall security.
\end{itemize}

\subsection{Example D: Design for Security Defense}
\label{sec:example-d}

This example represents papers that introduce new architectures and designs for preventing vulnerability exploitation. 
The following analysis is on the paper ``\textit{ZTD-Java: Mitigating Software Supply Chain Vulnerabilities via Zero-Trust Dependencies}'', which appeared at ICSE 2025~\cite{11029801}.
Papers with similar analyses may include~\cite{298068},~\cite{287304}, and~\cite{10646673}.

\subsubsection{Study Overview}

Modern applications rely heavily on third-party dependencies. As a result, they are susceptible to any reachable vulnerabilities within these libraries. Prior defenses were insufficient as they did not enforce zero-trust principles on the dependencies.
This paper defines and measures the effect of a Zero-Trust Architecture approach as applied to runtime dependencies.

\paragraph{\textbf{Problem Domain}}
Preventing exploitation of vulnerabilities in third-party dependencies.

\paragraph{\textbf{Methods}}
We applied two distinct methods in this study:
    \begin{itemize}
        \item \emph{Corpus Analysis:} This paper analyzed vulnerabilities in third-party libraries, popular third-party libraries, and a benchmark of real applications that use third-party libraries.
        \item \emph{Tool Evaluation}: This paper presented a system design and prototype for preventing vulnerability exploitation and evaluated its effectiveness, performance overhead, and the configuration effort it requires.
    \end{itemize}

\subsubsection{Stakeholder Identification}

\paragraph{Direct Stakeholders}
\begin{itemize}
    \item \textit{System operators}: Organizations deploying software that incorporates third-party dependencies. ZTD-Java provides them with the tool to protect their application from vulnerable dependencies.
    \item \textit{Software maintainers} - Maintainers of the vulnerable third-party libraries studied in the paper.
    \item \textit{The research team}: Authors of the work, who may face technical, professional, or reputational risks associated with proposing security defenses.
    \item \textit{Adversaries}: Malicious actors who attempt to exploit software applications through vulnerable dependencies.
\end{itemize}

\paragraph{Indirect Stakeholders}
\begin{itemize}
    \item \textit{Java maintainers}: Maintainers of the Java language and the Java Development Kit.
    \item \textit{End users}: Individuals and enterprises relying on software that uses third-party dependencies, whose security and privacy could be compromised by supply chain attacks.
    \item \textit{Broader public}: Society at large, whose services, security, and privacy can be impacted by large-scale software supply chain compromises.
\end{itemize}

\subsubsection{Ethical Considerations}

\begin{itemize}
    \item \textit{For maintainers of studied vulnerable libraries:} By studying vulnerabilities in third-party libraries, maintainers of these libraries may face additional reputational and legal risks.
    \item \textit{For adversaries:} Malicious actors can get motivation and insight from this paper to attack applications using vulnerable third-party libraries.
    \item \textit{For the research team:} The research team can face legal and reputational risks if the proposed security defenses are misconfigured or cause functional errors when deployed in applications.
\end{itemize}









\section{Discussion}
\label{sec:discussion}

Many papers and treatises have been written on the topic of ethics analysis in the context of cybersecurity.
We contribute some of our own thoughts here:
  next steps for research teams after stakeholder analysis (\cref{sec:discussion-ethics});
  whether ``doing no harm'' might lead to a chilling effect (\cref{sec:discussion-DoNoHarm});
  and the relation between ethics analysis and the standard sections on limitations and threats to validity (\cref{sec:discussion-EthicsVsLimitations}).

\subsection{From Stakeholder Identification to Ethics Analysis}
\label{sec:discussion-ethics}

This paper is intended to help research teams respond to the ethics requirements described in the USENIX Security 2026 Call for Papers~\cite{usenix-cfp-2025}, which mandates a stakeholder-based ethics analysis.  
By identifying stakeholder categories and providing method-informed guidance, we aim to support researchers in preparing a rigorous and transparent ethics section.  
Although our examples focus on stakeholder identification, we recognize that a complete ethics analysis will also involve reasoning about benefits, harms, and justifications for study design and publication.  
To that end, we encourage researchers to engage with relevant ethical frameworks.

Several traditions in moral philosophy can help guide reasoning about research ethics.  
A consequentialist approach evaluates actions based on outcomes, aiming to maximize benefit and minimize harm.  
A deontological approach emphasizes duties and rights, such as respecting consent and autonomy, even when outcomes are favorable.  
Virtue ethics focuses on the character and intentions of the researcher.  
These perspectives can reinforce each other or reveal tensions, especially when stakeholder interests conflict.  
We refer the reader to The Menlo Report and to recent work articulating how these frameworks apply in cybersecurity research~\cite{menlo-report, segal-trolley-2023}.

We further acknowledge that stakeholder identification is not a one-time process. 
Stakeholders are embedded within broader sociotechnical and institutional contexts, and their roles and relevance often emerge through their interactions with others.
Analyzing these interactions through the ethical analysis process can reveal additional stakeholders who may not be apparent when stakeholders are considered in isolation.
Thus, a comprehensive stakeholder identification process must occur iteratively with ethical analysis in order to accurately account for the stakeholder relationships.

We also note that ethical expectations may vary across institutional, national, and cultural contexts.  
Security researchers often work in multinational teams or study systems deployed globally.  
Therefore, it may be appropriate to frame stakeholder analysis within broader cultural values or legal systems.  
Prior work in engineering ethics and intercultural competence, including frameworks by Hofstede \etal~\cite{hofstede} and analyses by Zhu \etal~\cite{zhu2017engineering,zhu2020practicing}, argues that ethical decision-making is shaped by societal norms about authority, risk, and responsibility.

\subsection{Should Cybersecurity Researchers Do No Harm?}
\label{sec:discussion-DoNoHarm}

We reflect briefly on the assertion, implicit in the USENIX ethics policy, that cybersecurity research should not cause harm.  
This aligns with ethical norms in human subjects research and the broader computing community.  
Yet security research often involves adversarial contexts in which some stakeholder (\eg malicious actors) may be harmed by design.  
Moreover, some research may produce tools or knowledge that are dual-use.  
In such cases, the justification to proceed must be made cautiously and transparently, weighing harm to some against protection for others.  
The researcher’s role is not to avoid discomfort altogether, but to act with deliberation and integrity in anticipating and mitigating harms.

Why has this shift toward formalized ethics analysis occurred now?  
The recent USENIX Security policy likely responds, at least in part, to high-profile controversies in the field.  
One notable example is the retraction of the paper “On the Nature of Hypocritical Commits”~\cite{wu2021} from IEEE S\&P, which sparked sustained debate about consent, deception, and the ethical treatment of software developers.  
Such cases make clear that security research can cause real harm—and that ethical oversight is necessary to maintain trust within the community and with the public.

While we do not oppose ethics standards, we also caution the research community against too abruptly shifting away from adversarial research.  
Adversaries are at the heart of the discipline of cybersecurity.
Their capabilities are rapidly evolving, and there are active threats from both state-sponsored actors (\eg APT29 “Cozy Bear”~\cite{apt29}, China's PLA Unit 61398~\cite{unit61398}, the US NSA~\cite{tao-nsa}, and Israel's Unit 8200~\cite{unit8200}, to name but a few) and criminal syndicates operating via the dark web~\cite{darkweb}.  
These actors are not constrained by ethical review boards nor principles of research beneficence.  
If ethical mandates prevent researchers from exploring or disclosing certain risks, there is a real danger that defenders will be (further) outpaced by attackers.  
Ethical caution must be balanced with the imperative to understand, anticipate, and mitigate emerging threats.

In short: \textbf{\textit{These ethics mandates must not create a chilling effect on cybersecurity research}}.
Rather, it should enable researchers to proceed conscientiously, with awareness of the possible harms and a commitment to act with integrity.  
We hope our guide supports that confidence, both
  by helping researchers think carefully about whom their work may affect, and how;
  and
  by helping peer reviewers perform pragmatic assessments of harms to real stakeholders rather than overblown hypothetical ones.

\subsection{Ethics Analysis vs. Limitations and Threats to Validity}
\label{sec:discussion-EthicsVsLimitations}

Stakeholder ethics analysis is distinct from, but complementary to, the ``Limitations'' and ``Threats to Validity'' sections found in many computing research papers.  
Those sections typically address the extent to which the research findings are generalizable, robust, or methodologically sound.  
They are primarily epistemological in focus: concerned with what we can know from the results and how confidently we can make claims.  
In contrast, ethics analysis is \textit{normative}: it addresses what the researchers ought to do, what harms may arise, and what responsibilities they bear to others.  
Whereas a limitations section might admit that a study lacks ecological validity or statistical power, an ethics section should explain whether stakeholder interests were acknowledged and protected.  
Ethics analysis is structured not by research method alone but by the broader impact context of the work, and it is likely to include reflection on value-laden choices, such as whether to proceed with a study, disclose results, or disseminate findings responsibly.

\section{Conclusion}
\label{sec:conclusion}

Security research is conducted in a high-stakes, adversarial environment where ethical missteps can cause real harm—or prevent meaningful progress.  
In response to evolving expectations from the research community, we have presented a practical guide to stakeholder identification, grounded in empirical methods and accompanied by illustrative examples.  
Our goal is to help researchers anticipate who might be affected by their work and understand how methodological choices influence ethical exposure.

This guide may be used during project ideation, IRB preparation, or ethics section drafting.  
It may also assist reviewers and institutional reviewers in evaluating the completeness and clarity of submitted ethics analyses.  
Ultimately, we hope it enables thoughtful engagement with ethics without drifting into an overly restrictive ``do no harm'' mindset—one that might hinder critical inquiry rather than improving it.

\section*{Acknowledgment of Assistance}
\label{sec:acknowledgment}

This manuscript was developed with the assistance of various iterations of ChatGPT.
The human authors defined the structure and key ideas of the work.
ChatGPT was used to compose much of the initial draft, subject to human supervision and revision.


\clearpage

\bibliographystyle{IEEEtran}
\bibliography{bib/stakeholders}

\end{document}

%% file: misc/typesetting.tex
\usepackage{tcolorbox}
\usepackage[final, commandnameprefix=ifneeded]{changes} 

\newcommand{\eg}{\textit{e.g.},\xspace}
\newcommand{\etal}{\textit{et al.}\xspace}

\usepackage{amsthm}

\usepackage{enumitem}
\setlist[itemize]{leftmargin=*,noitemsep,topsep=0pt}
\setlist[enumerate]{leftmargin=*}

\usepackage{todonotes}
\usepackage{soul}

\ifDEBUG
    \newcommand{\DR}[1]{\todo[color=green,inline]{DR00:#1}}
    \newcommand{\JD}[1]{\todo[color=yellow,inline]{JD:#1}}
    \newcommand{\WJ}[1]{\todo[color=pink,inline]{WJ:#1}}

    \newcommand{\GKT}[1]{\todo[color=cyan,inline]{GKT:#1}}
    \newcommand{\KC}[1]{\todo[color=red,inline]{KC:#1}}

    \newcommand{\TODO}[1]{\hl{#1}}
\else
    \newcommand{\DR}[1]{}
    \newcommand{\JD}[1]{}
    \newcommand{\WJ}[1]{}
    \newcommand{\GKT}[1]{}
    \newcommand{\KC}[1]{}    
    \newcommand{\TODO}[1]{}
\fi

\usepackage{booktabs} 
\usepackage{colortbl}
\usepackage{amsmath}
\usepackage{amsfonts}
\usepackage{algpseudocode}
\usepackage{xspace}
\usepackage{multirow} 
\usepackage{balance} 
\usepackage{fancyvrb} 
\usepackage{url}
\usepackage{xcolor} 
\definecolor{table_gray}{gray}{0.9}

\usepackage{tikz} 
\usetikzlibrary{automata} 
\usetikzlibrary{positioning} 
\usetikzlibrary{arrows} 
\usetikzlibrary{calc} 
\usetikzlibrary{patterns} 
\usetikzlibrary{snakes} 

\usetikzlibrary{shapes.geometric, arrows.meta, positioning, fit}

\tikzstyle{process} = [rectangle, rounded corners=2pt, minimum width=3.8cm, minimum height=1.2cm, text centered, draw=black, fill=blue!10]
\tikzstyle{research} = [process, fill=gray!20]
\tikzstyle{ethics} = [process, fill=blue!10]
\tikzstyle{stop} = [regular polygon, regular polygon sides=8, draw=black, fill=red!30, minimum size=1.3cm]
\tikzstyle{decision} = [rectangle, draw=black, rounded corners=2pt, fill=red!10, minimum width=3.8cm, minimum height=1.2cm, text centered]
\tikzstyle{gate} = [diamond, aspect=2, text centered, draw=black, fill=yellow!20, minimum width=3.8cm]
\tikzstyle{arrow} = [thick, ->, >=stealth]
\tikzstyle{label} = [font=\small\itshape, align=center]
\tikzstyle{blocklabel} = [font=\footnotesize\bfseries]

\usepackage[group-separator={,}]{siunitx}

\VerbatimFootnotes 

\usepackage{cleveref}
\crefformat{section}{\S#2#1#3}
\crefname{figure}{Figure}{Figures}
\crefname{table}{Table}{Tables}
\crefname{listing}{Listing}{Listings}
\crefname{theorem}{Theorem}{Theorems}
\crefname{thm}{Theorem}{Theorems}
\crefname{lemma}{Lemma}{Lemmata}
\crefname{equation}{Eqt.}{Eqts.}
\crefformat{Grammar}{Grammar #1}

\usepackage{enumitem}
\usepackage{etoolbox}
\usepackage{hyperref}

\crefname{section}{Appendix}{Appendices}
\Crefname{section}{Appendix}{Appendices}

\makeatletter
\renewcommand\appendix{%
  \setcounter{section}{0}%
  \renewcommand\thesubsection{\arabic{subsection}}%
}
\makeatother

\usepackage{listings}

\usepackage{newfloat}
\usepackage{syntax}
\usepackage{framed}

\DeclareFloatingEnvironment[
  fileext   = logr,
  listname  = {List of Grammars},
  name      = Grammar,
  placement = htp
]{Grammar}

\usepackage[ruled,vlined]{algorithm2e}

\usepackage{nicefrac}

\usepackage{graphicx}
\usepackage{float}
\usepackage{multirow}

\usepackage{pifont} 
\usepackage{wasysym}

\usepackage{array}
\newcolumntype{P}[1]{>{\centering\arraybackslash}p{#1}}

\usepackage{graphicx}      
\usepackage{booktabs}      
\usepackage{caption}       

%% file: main.bbl
\begin{thebibliography}{10}
\providecommand{\url}[1]{#1}
\csname url@samestyle\endcsname
\providecommand{\newblock}{\relax}
\providecommand{\bibinfo}[2]{#2}
\providecommand{\BIBentrySTDinterwordspacing}{\spaceskip=0pt\relax}
\providecommand{\BIBentryALTinterwordstretchfactor}{4}
\providecommand{\BIBentryALTinterwordspacing}{\spaceskip=\fontdimen2\font plus
\BIBentryALTinterwordstretchfactor\fontdimen3\font minus \fontdimen4\font\relax}
\providecommand{\BIBforeignlanguage}[2]{{%
\expandafter\ifx\csname l@#1\endcsname\relax
\typeout{** WARNING: IEEEtran.bst: No hyphenation pattern has been}%
\typeout{** loaded for the language `#1'. Using the pattern for}%
\typeout{** the default language instead.}%
\else
\language=\csname l@#1\endcsname
\fi
#2}}
\providecommand{\BIBdecl}{\relax}
\BIBdecl

\bibitem{rogaway2015moral}
P.~Rogaway, ``The moral character of cryptographic work,'' \emph{Cryptology ePrint Archive}, 2015.

\bibitem{usenix-msg-2025}
\BIBentryALTinterwordspacing
{USENIX Security Symposium 2025 Program Co-Chairs}, ``Message from the usenix security ’25 program co-chairs,'' 2025, accessed: 2025-08-14. [Online]. Available: \url{https://www.usenix.org/sites/default/files/sec25_message.pdf}
\BIBentrySTDinterwordspacing

\bibitem{usenix-cfp-2026}
\BIBentryALTinterwordspacing
{USENIX Security Symposium 2026 Program Committee}, ``{USENIX Security ’26 Call for Papers},'' 2025, accessed: 2025-07-31. [Online]. Available: \url{https://www.usenix.org/conference/usenixsecurity26/call-for-papers}
\BIBentrySTDinterwordspacing

\bibitem{menlo-report}
\BIBentryALTinterwordspacing
D.~Dittrich and E.~Kenneally, ``The menlo report: Ethical principles guiding information and communication technology research,'' U.S. Department of Homeland Security, Tech. Rep., 2012. [Online]. Available: \url{https://www.caida.org/publications/papers/2012/menlo_report_actual_formatted/}
\BIBentrySTDinterwordspacing

\bibitem{beauchamp-principles}
T.~L. Beauchamp and J.~F. Childress, \emph{Principles of Biomedical Ethics}, 8th~ed.\hskip 1em plus 0.5em minus 0.4em\relax Oxford University Press, 2019.

\bibitem{segal-trolley-2023}
T.~Kohno, Y.~Acar, and W.~Loh, ``Ethical frameworks and computer security trolley problems: foundations for conversations,'' in \emph{2023 Proceedings of the 32th USENIX Security Symposium}, 2023, pp. 5145--5162.

\bibitem{friedman-2008-value}
\BIBentryALTinterwordspacing
B.~Friedman, P.~H. Kahn~Jr., and A.~Borning, \emph{Value Sensitive Design and Information Systems}.\hskip 1em plus 0.5em minus 0.4em\relax John Wiley \& Sons, Ltd, 2008, ch.~4, pp. 69--101. [Online]. Available: \url{https://onlinelibrary.wiley.com/doi/abs/10.1002/9780470281819.ch4}
\BIBentrySTDinterwordspacing

\bibitem{mill-utilitarianism}
J.~S. Mill, \emph{Utilitarianism}.\hskip 1em plus 0.5em minus 0.4em\relax Parker, Son, and Bourn, 1863.

\bibitem{kant-metaphysics}
I.~Kant, \emph{Groundwork of the Metaphysics of Morals}, 1785, translated by Mary Gregor (Cambridge Unviersity Press, 2nd ed., 2012).

\bibitem{anscombe-modern-moral-philosophy}
G.~Anscombe, ``Modern moral philosophy,'' \emph{Philosophy}, vol.~33, no. 124, pp. 1--19, 1958.

\bibitem{shostack2014threat}
A.~Shostack, \emph{Threat Modeling: Designing for Security}.\hskip 1em plus 0.5em minus 0.4em\relax Wiley, 2014.

\bibitem{795198}
H.~Sharp, A.~Finkelstein, and G.~Galal, ``Stakeholder identification in the requirements engineering process,'' in \emph{Proceedings. Tenth International Workshop on Database and Expert Systems Applications. DEXA 99}, 1999, pp. 387--391.

\bibitem{freeman1984}
R.~E. Freeman, \emph{Strategic Management: A Stakeholder Approach}.\hskip 1em plus 0.5em minus 0.4em\relax Pitman Publishing, 1984.

\bibitem{leveson2011engineering}
N.~G. Leveson, \emph{Engineering a Safer World: Systems Thinking Applied to Safety}.\hskip 1em plus 0.5em minus 0.4em\relax MIT Press, 2012.

\bibitem{sommerville1997guide}
I.~Sommerville and P.~Sawyer, \emph{Requirements Engineering: A Good Practice Guide}, 1st~ed.\hskip 1em plus 0.5em minus 0.4em\relax USA: John Wiley \& Sons, Inc., 1997.

\bibitem{sommerville1997viewpoints}
------, ``Viewpoints: principles, problems and a practical approach to requirements engineering,'' \emph{Annals of software engineering}, vol.~3, no.~1, pp. 101--130, 1997.

\bibitem{nist-glossary}
\BIBentryALTinterwordspacing
{National Institute of Standards and Technology}, ``Computer security resource center glossary: Cybersecurity,'' 2024, accessed: 2025-07-31. [Online]. Available: \url{https://csrc.nist.gov/glossary/term/cybersecurity}
\BIBentrySTDinterwordspacing

\bibitem{sp-cfp-2025}
\BIBentryALTinterwordspacing
{IEEE S\&P 2025 Program Committee}, ``{Call for Papers},'' 2024, accessed: 2025-07-31. [Online]. Available: \url{https://sp2025.ieee-security.org/cfpapers.html}
\BIBentrySTDinterwordspacing

\bibitem{usenix-cfp-2025}
\BIBentryALTinterwordspacing
{USENIX Security Symposium 2025 Program Committee}, ``{USENIX Security '25 Call for Papers},'' 2024, accessed: 2025-07-31. [Online]. Available: \url{https://www.usenix.org/conference/usenixsecurity25/call-for-papers}
\BIBentrySTDinterwordspacing

\bibitem{ndss-cfp-2025}
\BIBentryALTinterwordspacing
{NDSS Symposium 2025 Program Committee}, ``{NDSS Symposium 2025 Call for Papers},'' 2024, accessed: 2025-07-31. [Online]. Available: \url{https://www.ndss-symposium.org/ndss2025/submissions/call-for-papers/}
\BIBentrySTDinterwordspacing

\bibitem{ccs-cfp-2025}
\BIBentryALTinterwordspacing
{ACM CCS 2025 Program Committee}, ``{Call for Papers},'' 2024, accessed: 2025-07-31. [Online]. Available: \url{https://www.sigsac.org/ccs/CCS2025/call-for-papers/}
\BIBentrySTDinterwordspacing

\bibitem{narayanan2008robust}
A.~Narayanan and V.~Shmatikov, ``Robust de-anonymization of large sparse datasets,'' in \emph{Proceedings of the IEEE Symposium on Security and Privacy}.\hskip 1em plus 0.5em minus 0.4em\relax IEEE, 2008, pp. 111--125.

\bibitem{metcalf2016ethics}
J.~Metcalf, E.~Keller, and d.~boyd, ``Ethics in emerging technology: A particular focus on data and privacy,'' \emph{Journal of Information, Communication and Ethics in Society}, vol.~14, no.~2, pp. 77--92, 2016.

\bibitem{brumley2008automatic}
D.~Brumley, P.~Poosankam, D.~Song, and J.~Zheng, ``Automatic patch-based exploit generation is possible: Techniques and implications,'' in \emph{2008 IEEE Symposium on Security and Privacy (sp 2008)}.\hskip 1em plus 0.5em minus 0.4em\relax IEEE, 2008, pp. 143--157.

\bibitem{riebe2019dual}
T.~Riebe and C.~Reuter, ``Dual-use and dilemmas for cybersecurity, peace and technology assessment,'' in \emph{Information Technology for Peace and Security: IT Applications and Infrastructures in Conflicts, Crises, War, and Peace}.\hskip 1em plus 0.5em minus 0.4em\relax Springer, 2019, pp. 165--183.

\bibitem{mutlu2019rowhammer}
O.~Mutlu and J.~S. Kim, ``Rowhammer: A retrospective,'' \emph{IEEE Transactions on Computer-Aided Design of Integrated Circuits and Systems}, vol.~39, no.~8, pp. 1555--1571, 2019.

\bibitem{sigsoft-standards}
{P. Ralph et al.}, ``{Empirical Standards for Software Engineering Research},'' \emph{arXiv:2010.03525 [cs.SE]}, 2021.

\bibitem{kalu2024industry}
K.~G. Kalu, T.~Singla, C.~Okafor, S.~Torres-Arias, and J.~C. Davis, ``An industry interview study of software signing for supply chain security,'' in \emph{2025 Proceedings of the 34th USENIX Security Symposium}, 2025, pp. 81--100.

\bibitem{285411}
\BIBentryALTinterwordspacing
R.~Ramesh, A.~Vyas, and R.~Ensafi, ``"all of them claim to be the best": Multi-perspective study of {VPN} users and {VPN} providers,'' in \emph{32nd USENIX Security Symposium (USENIX Security 23)}.\hskip 1em plus 0.5em minus 0.4em\relax Anaheim, CA: USENIX Association, Aug. 2023, pp. 5773--5789. [Online]. Available: \url{https://www.usenix.org/conference/usenixsecurity23/presentation/ramesh-vpn}
\BIBentrySTDinterwordspacing

\bibitem{airtag}
J.~Chen, X.~Ma, L.~Luo, and Q.~Zeng, ``Tracking you from a thousand miles away! turning a bluetooth device into an apple airtag without root privileges,'' in \emph{2025 Proceedings of the 34th USENIX Security Symposium}, 2025, pp. 4345--4362.

\bibitem{dalle}
C.~Villa, S.~Mirza, and C.~Pöpper, ``Exposing the guardrails: Reverse-engineering and jailbreaking safety filters in dall·e text-to-image pipelines,'' in \emph{2025 Proceedings of the 34th USENIX Security Symposium}, 2025, pp. 897--916.

\bibitem{10298483}
P.~C. Amusuo, R.~A.~C. Méndez, Z.~Xu, A.~Machiry, and J.~C. Davis, ``Systematically detecting packet validation vulnerabilities in embedded network stacks,'' in \emph{2023 38th IEEE/ACM International Conference on Automated Software Engineering (ASE)}, 2023, pp. 926--938.

\bibitem{catch22}
C.~Munteanu, G.~Smaragdakis, A.~Feldmann, and T.~Fiebig, ``Catch-22: Uncovering compromised hosts using ssh public keys,'' in \emph{2025 Proceedings of the 34th USENIX Security Symposium}, 2025, pp. 861--878.

\bibitem{amusuo2025unit}
P.~C. Amusuo, O.~Cochell, T.~L. Lievre, P.~V. Patil, A.~Machiry, and J.~C. Davis, ``Do unit proofs work? an empirical study of compositional bounded model checking for memory safety verification,'' \emph{arXiv preprint arXiv:2503.13762}, 2025.

\bibitem{298150}
\BIBentryALTinterwordspacing
B.~Kondracki and N.~Nikiforakis, ``Smudged fingerprints: Characterizing and improving the performance of web application fingerprinting,'' in \emph{33rd USENIX Security Symposium (USENIX Security 24)}.\hskip 1em plus 0.5em minus 0.4em\relax Philadelphia, PA: USENIX Association, Aug. 2024, pp. 4625--4640. [Online]. Available: \url{https://www.usenix.org/conference/usenixsecurity24/presentation/kondracki}
\BIBentrySTDinterwordspacing

\bibitem{10646663}
S.~Ullah, M.~Han, S.~Pujar, H.~Pearce, A.~Coskun, and G.~Stringhini, ``Llms cannot reliably identify and reason about security vulnerabilities (yet?): A comprehensive evaluation, framework, and benchmarks,'' in \emph{2024 IEEE Symposium on Security and Privacy (SP)}, 2024, pp. 862--880.

\bibitem{11029801}
P.~C. Amusuo, K.~A. Robinson, T.~Singla, H.~Peng, A.~Machiry, S.~Torres-Arias, L.~Simon, and J.~C. Davis, ``{ZTD}$_{\text{{j}ava}}$: Mitigating software supply chain vulnerabilities via zero-trust dependencies,'' in \emph{2025 IEEE/ACM 47th International Conference on Software Engineering (ICSE)}, 2025, pp. 1294--1306.

\bibitem{298186}
\BIBentryALTinterwordspacing
J.~Zhang, J.~Huang, L.~Zhao, D.~Chen, and {\c C}.~K. Ko{\c c}, ``{ENG25519}: Faster {TLS} 1.3 handshake using optimized x25519 and ed25519,'' in \emph{33rd USENIX Security Symposium (USENIX Security 24)}.\hskip 1em plus 0.5em minus 0.4em\relax Philadelphia, PA: USENIX Association, Aug. 2024, pp. 6381--6398. [Online]. Available: \url{https://www.usenix.org/conference/usenixsecurity24/presentation/zhang-jipeng}
\BIBentrySTDinterwordspacing

\bibitem{285379}
\BIBentryALTinterwordspacing
A.~Hilton, C.~Deccio, and J.~Davis, ``Fourteen years in the life: A root {Server{\textquoteright}s} perspective on {DNS} resolver security,'' in \emph{32nd USENIX Security Symposium (USENIX Security 23)}.\hskip 1em plus 0.5em minus 0.4em\relax Anaheim, CA: USENIX Association, Aug. 2023, pp. 3171--3186. [Online]. Available: \url{https://www.usenix.org/conference/usenixsecurity23/presentation/hilton}
\BIBentrySTDinterwordspacing

\bibitem{11023368}
T.~Marjanov and A.~Hutchings, ``Sok: Digging into the digital underworld of stolen data markets,'' in \emph{2025 IEEE Symposium on Security and Privacy (SP)}, 2025, pp. 1--18.

\bibitem{11023489}
K.~Beadle, K.~I. Turk, A.~Eusebi, M.~Tran, M.~Ordekian, E.~Mariconti, Y.~Zou, and M.~Vasek, ``Sok: A privacy framework for security research using social media data,'' in \emph{2025 IEEE Symposium on Security and Privacy (SP)}, 2025, pp. 1178--1196.

\bibitem{10646648}
A.~Augusto, R.~Belchior, M.~Correia, A.~Vasconcelos, L.~Zhang, and T.~Hardjono, ``Sok: Security and privacy of blockchain interoperability,'' in \emph{2024 IEEE Symposium on Security and Privacy (SP)}, 2024, pp. 3840--3865.

\bibitem{zksnarks}
J.~Liang, D.~Hu, P.~Wu, Y.~Yang, Q.~Shen, and Z.~Wu, ``Sok: Understanding zk-snarks: The gap between research and practice,'' in \emph{2025 Proceedings of the 34th USENIX Security Symposium}, 2025, pp. 2085--2104.

\bibitem{287290}
\BIBentryALTinterwordspacing
S.~Hebrok, S.~Nachtigall, M.~Maehren, N.~Erinola, R.~Merget, J.~Somorovsky, and J.~Schwenk, ``We really need to talk about session tickets: A {Large-Scale} analysis of cryptographic dangers with {TLS} session tickets,'' in \emph{32nd USENIX Security Symposium (USENIX Security 23)}.\hskip 1em plus 0.5em minus 0.4em\relax Anaheim, CA: USENIX Association, Aug. 2023, pp. 4877--4894. [Online]. Available: \url{https://www.usenix.org/conference/usenixsecurity23/presentation/hebrok}
\BIBentrySTDinterwordspacing

\bibitem{10646755}
E.~Wang, J.~Chen, W.~Xie, C.~Wang, Y.~Gao, Z.~Wang, H.~Duan, Y.~Liu, and B.~Wang, ``Where urls become weapons: Automated discovery of ssrf vulnerabilities in web applications,'' in \emph{2024 IEEE Symposium on Security and Privacy (SP)}, 2024, pp. 239--257.

\bibitem{298202}
\BIBentryALTinterwordspacing
M.~Busch, P.~Mao, and M.~Payer, ``Spill the {TeA}: An empirical study of trusted application rollback prevention on android smartphones,'' in \emph{33rd USENIX Security Symposium (USENIX Security 24)}.\hskip 1em plus 0.5em minus 0.4em\relax Philadelphia, PA: USENIX Association, Aug. 2024, pp. 5071--5088. [Online]. Available: \url{https://www.usenix.org/conference/usenixsecurity24/presentation/busch-tea}
\BIBentrySTDinterwordspacing

\bibitem{298148}
\BIBentryALTinterwordspacing
G.~Calderonio, M.~M. Ali, and J.~Polakis, ``Fledging will continue until privacy improves: Empirical analysis of google{\textquoteright}s {Privacy-Preserving} targeted advertising,'' in \emph{33rd USENIX Security Symposium (USENIX Security 24)}.\hskip 1em plus 0.5em minus 0.4em\relax Philadelphia, PA: USENIX Association, Aug. 2024, pp. 4121--4138. [Online]. Available: \url{https://www.usenix.org/conference/usenixsecurity24/presentation/calderonio}
\BIBentrySTDinterwordspacing

\bibitem{285517}
\BIBentryALTinterwordspacing
K.~L. Wu, M.~H. Hue, N.~M. Poon, K.~M. Leung, W.~Y. Po, K.~T. Wong, S.~H. Hui, and S.~Y. Chau, ``Back to school: On the ({In)Security} of academic {VPNs},'' in \emph{32nd USENIX Security Symposium (USENIX Security 23)}.\hskip 1em plus 0.5em minus 0.4em\relax Anaheim, CA: USENIX Association, Aug. 2023, pp. 5737--5754. [Online]. Available: \url{https://www.usenix.org/conference/usenixsecurity23/presentation/wu-ka-lok}
\BIBentrySTDinterwordspacing

\bibitem{11023474}
G.~t. Napel, M.~van Eeten, and S.~Parkin, ``Speedrunning the maze: Meeting regulatory patching deadlines in a large enterprise environment,'' in \emph{2025 IEEE Symposium on Security and Privacy (SP)}, 2025, pp. 504--521.

\bibitem{298068}
\BIBentryALTinterwordspacing
Y.~Chen, Q.~Yin, Q.~Li, Z.~Liu, K.~Xu, Y.~Xu, M.~Xu, Z.~Liu, and J.~Wu, ``Learning with semantics: Towards a {Semantics-Aware} routing anomaly detection system,'' in \emph{33rd USENIX Security Symposium (USENIX Security 24)}.\hskip 1em plus 0.5em minus 0.4em\relax Philadelphia, PA: USENIX Association, Aug. 2024, pp. 5143--5160. [Online]. Available: \url{https://www.usenix.org/conference/usenixsecurity24/presentation/chen-yihao}
\BIBentrySTDinterwordspacing

\bibitem{287304}
\BIBentryALTinterwordspacing
T.~Wallez, J.~Protzenko, B.~Beurdouche, and K.~Bhargavan, ``{TreeSync}: Authenticated group management for messaging layer security,'' in \emph{32nd USENIX Security Symposium (USENIX Security 23)}.\hskip 1em plus 0.5em minus 0.4em\relax Anaheim, CA: USENIX Association, Aug. 2023, pp. 1217--1233. [Online]. Available: \url{https://www.usenix.org/conference/usenixsecurity23/presentation/wallez}
\BIBentrySTDinterwordspacing

\bibitem{10646673}
Z.~Cheng, Q.~Lv, J.~Liang, Y.~Wang, D.~Sun, T.~Pasquier, and X.~Han, ``Kairos: Practical intrusion detection and investigation using whole-system provenance,'' in \emph{2024 IEEE Symposium on Security and Privacy (SP)}, 2024, pp. 3533--3551.

\bibitem{hofstede}
G.~Hofstede, G.~J. Hofstede, and M.~Minkov, \emph{Cultures and Organizations: Software of the Mind}, 3rd~ed.\hskip 1em plus 0.5em minus 0.4em\relax McGraw-Hill, 2010.

\bibitem{zhu2017engineering}
Q.~Zhu and B.~K. Jesiek, ``Engineering ethics in global context: Four fundamental approaches,'' in \emph{2017 ASEE Annual Conference \& Exposition}, 2017.

\bibitem{zhu2020practicing}
------, ``Practicing engineering ethics in global context: A comparative study of expert and novice approaches to cross-cultural ethical situations,'' \emph{Science and Engineering Ethics}, vol.~26, no.~4, pp. 2097--2120, 2020.

\bibitem{wu2021}
Y.~Wu and K.~Lu, ``On the feasibility of stealthily introducing vulnerabilities in open-source software via hypocrite commit,'' in \emph{2021 IEEE Symposium on Security and Privacy (SP)}, 2021, retracted.

\bibitem{apt29}
\BIBentryALTinterwordspacing
{Cybersecurity and Infrastructure Security Agency}, ``{Russian Foreign Intelligence Service (SVR) Exploiting JetBrains TeamCity CVE Globally},'' 2023, accessed: 2025-07-31. [Online]. Available: \url{https://www.cisa.gov/news-events/cybersecurity-advisories/aa23-347a}
\BIBentrySTDinterwordspacing

\bibitem{unit61398}
\BIBentryALTinterwordspacing
D.~Mcwhorter, ``{Mandiant Exposes APT1 – One of China's Cyber Espionage Units – and Releases 3,000 Indicators},'' 2013, accessed: 2025-07-31. [Online]. Available: \url{https://www.mandiant.com/resources/apt1-exposing-one-of-chinas-cyber-espionage-units}
\BIBentrySTDinterwordspacing

\bibitem{tao-nsa}
\BIBentryALTinterwordspacing
{Von SPIEGEL Staff}, ``{Documents Reveal Top NSA Hacking Unit},'' 2013, accessed: 2025-07-31. [Online]. Available: \url{https://www.spiegel.de/international/world/a-940969.html}
\BIBentrySTDinterwordspacing

\bibitem{unit8200}
\BIBentryALTinterwordspacing
``Unit 8200,'' accessed: 2025-08-19. [Online]. Available: \url{https://en.wikipedia.org/wiki/Unit_8200}
\BIBentrySTDinterwordspacing

\bibitem{darkweb}
\BIBentryALTinterwordspacing
``Internet organised crime threat assessment (iocta),'' Europol, Tech. Rep., 2023. [Online]. Available: \url{https://www.europol.europa.eu/cms/sites/default/files/documents/IOCTA%202023%20-%20EN_0.pdf}
\BIBentrySTDinterwordspacing

\end{thebibliography}
